\documentclass[sn-mathphys-num]{sn-jnl}

\usepackage{graphicx}%
\usepackage{multirow}%
\usepackage{amsmath,amssymb,amsfonts}%
\usepackage{amsthm}%
\usepackage{mathrsfs}%
\usepackage[title]{appendix}%
\usepackage{xcolor}%
\usepackage{textcomp}%
\usepackage{manyfoot}%
\usepackage{booktabs}%
\usepackage{algorithm}%
\usepackage{algorithmicx}%
\usepackage{algpseudocode}%
\usepackage{listings}%

\theoremstyle{thmstyleone}%
\theoremstyle{thmstyletwo}%

\theoremstyle{thmstylethree}%

\begin{document}

\title[Article Title]{Perspective: Floquet engineering topological states from effective models towards realistic materials}

\author[1,2]{\fnm{Fangyang} \sur{Zhan}}
\equalcont{These authors contributed equally to this work.}

\author[3]{\fnm{Rui} \sur{Chen}}
\equalcont{These authors contributed equally to this work.}

\author[1,2]{\fnm{Zhen} \sur{Ning}}

\author[1,2]{\fnm{Da-Shuai} \sur{Ma}}
\author[1,2]{\fnm{Ziming} \sur{Wang}}


\author*[1,2]{\fnm{Dong-Hui} \sur{Xu}}\email{donghuixu@cqu.edu.cn}

\author*[1,2]{\fnm{Rui} \sur{Wang}}\email{rcwang@cqu.edu.cn}

\affil[1]{\orgdiv{Institute for Structure and Function $\&$ Department of Physics $\&$ Chongqing Key Laboratory for Strongly Coupled Physics}, \orgname{Chongqing University}, \orgaddress{\city{Chongqing}, \postcode{400044}, \country{P. R. China}}}

\affil[2]{\orgdiv{Center of Quantum Materials and Devices}, \orgname{Chongqing University}, \orgaddress{\city{Chongqing}, \postcode{400044}, \country{P. R. China}}}

\affil[3]{\orgdiv{Department of Physics}, \orgname{Hubei University}, \orgaddress{\city{Wuhan}, \postcode{430062}, \country{P. R. China}}}



\abstract{With significant advances in classifying and cataloguing topological matter, the focus of topological physics has shifted towards quantum control, particularly the creation and manipulation of topological phases of matter. Floquet engineering, the concept of tailoring a system by periodic fields, offers a powerful tool to manipulate electronic properties of condensed systems, and even to create exotic non-equilibrium topological states that are impossibly present in equilibrium scenarios. In this perspective, we give a brief review of recent progress in theoretical investigations of Floquet engineering topological states from effective models towards realistic materials. We show that light irradiation can realize various desired topological states through the introduction of symmetry breaking, such as first- and higher-order Weyl fermions, quadrupole topological insulator with periodic driving and disorder, quantum anomalous Hall effects with a tunable Chern number, as well as beyond. Moreover, based on first-principles calculations and Floquet theorem, we show several realistic material candidates proposed as potential hosts for promising Floquet topological states, facilitating their verification in experiments. We believe that our perspective on Floquet engineering of topological states will advance further studies of rich exotic light-induced phenomena in condensed matter physics.}





\maketitle

\section{Introduction}\label{sec1}
In the past decades, the study of topological phases of matter has been a significant subject~\cite{RevModPhys.82.3045,RevModPhys.83.1057,RevModPhys.90.015001,RevModPhys.88.021004,RevModPhys.93.025002,PhysRevB.83.205101}. Due to their profound significance and extensive potential for next-generation devices with ultralow power dissipation, topological materials have emerged as a frontier in fields of condensed matter physics as well as material science. Recent research has seen a surge in classifying and realizing topological states in crystalline materials, thanks to progress in symmetry-based theoretical frameworks and computational methods~\cite{RevModPhys.88.035005,PhysRevB.55.1142,1.3149495,PhysRevB.78.195125,2022symme,PhysRevX.7.041069,spacegroup2013}. Various catalogues of topological materials, such as nonmagnetic and magnetic topological electronic materials \cite{2019Vergniory,2019Catalogue,2019Comprehensive,science.abg9094,Xu2020,2022Regnault,2022Progress}, topological phononic materials~\cite{science.adf8458,adfm.201904784,2021LiComput,PhysRevLett.120.016401,PhysRevLett.123.065501,PhysRevLett.124.105303,PhysRevLett.126.185301}, and topological superconductors~\cite{1093nwaa169,PhysRevResearch.2.013064,2019Variants,sciadv.aaz8367,PhysRevB.101.245128,PhysRevResearch.3.023086,PhysRevX.12.011021,PhysRevResearch.3.013243,PhysRevLett.129.027001}, have been successively established. More recently, by further utilizing symmetry arguments, researchers have comprehensively constructed the effective models of all magnetic space groups~\cite{sciadv.aat8685,2019Comprehensive,PhysRevB.104.085137,PhysRevB.105.155156} and established the encyclopedia of emergent quasiparticles in three-dimensional crystals~\cite{bradlyn2016beyond,YU2022375}, strongly facilitating the development of topological physics. However, the current focus mainly lies in investigating equilibrium states of topological physics, while studies on topological physics in non-equilibrium scenarios are still in the infancy. With the increasing achievements of topological states and topological materials, performing investigations of quantum control such as creation and manipulation of topological states is an imperative task.

Light-matter interaction is an important approach to dynamically modulate material properties on ultrafast timescales, enabling the creation of exotic non-equilibrium topological states that are otherwise not possible in the equilibrium cases~\cite{PhysRevB.99.214302,PhysRevResearch.2.043408,Light2022,Oka1,Oka2,lindner2011floquet,annurev013423,2020nonequilibrium,RevModPhys.93.041002}. Among various mechanisms of light-matter interaction, of particular interest is the concept of Floquet engineering. Within the framework of Floquet theory~\cite{PhysRev.138.B979,PhysRevB.34.3625,PhysRevA.7.2203,Gesztesy1981JPA}, periodic light field transfers Bloch energy bands of crystalline solids to periodic Floquet-Bloch sidebands through multiphoton absorption or emission. Thus, light driving offers a mean to manipulate electronic structures, enabling a great potential to control electronic topology out of equilibrium in materials. By appropriately selecting incident light that matches the target system, Floquet engineering can provide a wide range of pathways for dynamically manipulating topological states and even inducing topological phase transitions. For instance, light can gap out the Dirac cone in graphene or drive a band inversion in semiconductor quantum wells, thereby developing the notion of a Floquet topological insulator (FTI)~\cite{lindner2011floquet,pssr.201206451,PhysRevB.93.045121,PhysRevB.106.224306,PhysRevX.6.041001,PhysRevLett.110.016802}. The  topologically nontrivial band gap can also be induced from avoided crossings of photon-dressed Floquet sidebands via the optical Stark effect~\cite{2015Valley,PhysRevB.97.045307,nanolett.6b04419,science.aal2241}.

The Floquet topological phases are strongly dependent on the drive frequency, amplitude, and polarization of incident light, exhibiting high tunability. It has been demonstrated that irradiation of circularly polarized light (CPL) can break the time-reversal ($\mathcal{T}$) symmetry, and thus quantum anomalous Hall (QAH) insulators with a non-zero Chern number can be obtained under irradiation of CPL~\cite{Oka1,Inoue,Oka2,Ezawa,Mimp,PhysRevLett.120.156406,PhysRevB.105.035103}. Unlike the QAH effect that is present in magnetic materials, the Floquet QAH effect does not require initial magnetism and can generally be present in light-irradiated magnetic and nonmagnetic materials~\cite{PhysRevB.105.L081115,nanolett.2c04651}. Besides, Floquet engineering can give rise to Chern flat bands with tunable large Chern numbers in twisted systems irradiated by CPL~\cite{PhysRevResearch.1.023031,PhysRevB.103.195146,PhysRevB.103.014310,PhysRevResearch.2.033494}. Inspired by the development of conventional FTI with the dipole polarization, it is found that periodic driving can further lead to Floquet higher-order topological insulators with multipole polarization~\cite{Ghosh_2024,PhysRevB.105.115418,PhysRevB.103.L041402,PhysRevB.99.045441,PhysRevB.100.115403,PhysRevLett.123.016806,PhysRevB.101.235403,PhysRevLett.124.057001,PhysRevLett.124.216601,PhysRevB.105.L201114}. Beyond gapped topological phases, tailoring topological semimetallic phases with first-order or higher-order topology through light irradiation have also been intensely studied~\cite{XLDU2022,Chen2018PRBLNSM,ZMWangPRB,Wang_2014}. Through changing the propagation or polarization direction of incident light, light control of symmetry breaking can be accordingly achieved, and periodic driving via light irradiation offers a fascinating avenue to realize desired gapless topological fermions. The typical examples are the Floquet Weyl semimetals (WSMs) with highly controllable Weyl nodes, which can be generated in light-irradiated topological insulators~\cite{Wang_2014}, Dirac semimetals (DSMs)~\cite{PhysRevLett.128.066602,PhysRevB.99.075121,hubener2017,PhysRevB.94.081103,PhysRevB.94.121106}, nodal-line semimetals (NLSMs)~\cite{PhysRevLett.117.087402,PhysRevB.94.041409,PhysRevB.102.201105,PhysRevB.96.041205}.

On the other hand, disorder plays an important role on the observation of topological edge or surface transport in a realistic topological system~\cite{TAI-1,TAI-2,3DTAI,expTAI,PTAI,Titum1,Titum2}, and even can induce a phase transition from a topologically trivial insulator to a topological Anderson insulator (TAI) phase~\cite{TAI-1}. In periodically driven systems, it has been demonstrated that disorder can induce topological phases that go beyond the well established paradigm of static disorder-induced topological phases~\cite{Titum1,Titum2,wauters2019prl,arxiv03727}. The topological phase arises from the interplay of disorder and periodic driving, dubbed a Floquet TAI. Experimentally, the Floquet topological sates as well as their emerged exotic properties can be captured by time-resolved and angle-resolved photoemission spectroscopy (TrARPES) or time-resolved transport measurements~\cite{2022arpes,science.1239834,2016Selective,PhysRevX.10.041013,nanolett.1c00801,choi2024,merboldt2024}. For instance, the Floquet sidebands and CPL-induced gap of topological surface states were observed in three-dimensional topological insulator Bi$_2$Se$_3$ by using TrARPES~\cite{2016Selective,science.1239834}. Through the application of ultrafast time-resolved transport measurements using a laser-triggered ultrafast photoconductive switch, the light-induced anomalous Hall effect was confirmed in CPL-driven graphene~\cite{mciver2020light}. Recently, the Floquet band engineering in a semiconductor has achieved important progress in experiments, such as pseudospin-selective Floquet sidebands in black phosphorus~\cite{2023Pseudospin}, and optical control of valley polarization in semiconductors MoS$_2$~\cite{2024valleytronics} and BN~\cite{2024BN}.

Overall, design and control of topological states via Floquet engineering has gradually become an attractive focus in topological physics, with rich exotic phenomena and promising application prospects. To date, numerous representative works have been proposed theoretically, suggesting rich light-induced topological phases based on effective modes, thereby significantly advancing studies of Floquet engineering in condensed-matter community~\cite{Light2022,annurev013423,2020nonequilibrium,2018Theoretical,Giovannini_2020,RODRIGUEZVEGA2021168434,LIU2023100705}. However, only a few experimental evidences have confirmed light-induced topological sates and phase transitions in periodically driven systems~\cite{nanolett.1c00801,2016Selective,science.1239834,mciver2020light,PhysRevX.10.041013}. The exploration of realistic material candidates that can realize these theoretical effective models is relatively slow. Therefore, to drive Floquet engineering forward by laying down a foundation for experiments, the combination of first-principles calculations and Floquet theorem is an effective mean to predict novel topological properties in realistic material candidates. In comparison with effective model calculations, the first-principles based approach can map momentum- and spin- resolved Floquet-Bloch bands in whole Brillouin zone of solids, and further depict complex and entangled band manifolds on ultrafast timescales~\cite{hubener2017,PhysRevB.99.075121,PhysRevB.102.201105,PhysRevLett.120.156406,LIU2023100705,PhysRevB.105.L081115,nanolett.2c04651,PhysRevLett.120.237403,PhysRevB.107.085151,advs.202101508}. Therefore, this perspective will provide a brief review of recent progress in theoretical investigations of Floquet engineering topological states from effective models towards realistic materials. In particular, most of material candidates are well studied in literature, facilitating the realization of Floquet engineering topological states and their device design in experiments.

\section{Basic method of Floquet engineering}\label{sec2}
In this section, we review theoretical formalisms and computational methods for Floquet engineering electronic states in crystalline materials under irradiation of time-periodic light fields.

\subsection{Floquet-Bloch Hamiltonian}\label{subsec21}
Firstly, we consider a system driven by a time-periodic and space-homogeneous light field with a Bloch Hamiltonian $H(\mathbf{k})$ with crystal momentum $\mathbf{k}$. The light field can be described as a vector potential as $\mathbf{A}(t)=\mathbf{A}(t+T)$ with period $T=2\pi/\omega$ and the light frequency $\omega$, and then the polarized electric field is $\mathbf{E}(t)=-\partial_{t}\mathbf{A}(t)$. In the presence of periodic drive, we obtain a time-dependent Hamiltonian $H(\mathbf{k},t)=H[\mathbf{k}+\mathbf{A}(t)]$. When the Hamiltonian is time periodic, the Floquet theorem~\cite{PhysRev.138.B979,PhysRevB.34.3625,PhysRevA.7.2203,Gesztesy1981JPA} allows us to map it to a time-independent Hamiltonian.
Specially, the Bloch wavefunction $|\Psi(\mathbf{k})\rangle$ develops into the time-dependent formalism as
\begin{equation}
|\Psi(\mathbf{k},t)\rangle=\exp{[-i\varepsilon(\mathbf{k})t]}|\Phi(\mathbf{k},t)\rangle
\end{equation}
with time-periodic auxiliary function $|\Phi(\mathbf{k},t)\rangle=|\Phi(\mathbf{k},t+T)\rangle$, which can be expanded in discrete Fourier series as
\begin{equation}\label{WV1}
|\Phi(\mathbf{k},t)\rangle=\Sigma_\alpha e^{-\alpha\omega t}|u^{\alpha}(\mathbf{k})\rangle,
\end{equation}
where $\alpha\in(-\infty,+\infty)$ is an integer and termed as the Floquet index. Besides, the electronic wavefunction $|\Psi(\mathbf{k},t)\rangle$ in light-driven
crystals is determined through the time-dependent Schr\"{o}dinger equation as
\begin{equation}\label{Ham2}
i\frac{\partial}{\partial t}|\Psi(\mathbf{k},t)\rangle=H(\mathbf{k},t)|\Psi(\mathbf{k},t)\rangle.
\end{equation}
Combining with Eqs. (\ref{WV1}) and (\ref{Ham2}), the time-dependent Schr\"{o}dinger equation can be transformed into a series of time-independent equations as
\begin{equation}
\Sigma_{\alpha}H^{mn}(\mathbf{k})|u^{\alpha}(\mathbf{k})\rangle=[\varepsilon(\mathbf{k})+\beta\hbar\omega]|u^{\beta}(\mathbf{k})\rangle,
\end{equation}
with
\begin{equation}
H^{\alpha-\beta}(\mathbf{k})=\frac{1}{T}\int_{0}^{T}H(\mathbf{k},t)e^{-i(\alpha-\beta)\omega t}dt,
\end{equation}
Here, the time-independent square matrix $H^{(\alpha-\beta)}(\mathbf{k})$ is dubbed as Floquet-Bloch Hamiltonian. With this representation, the time-dependent
Schr\"{o}dinger equation is mapped to an eigenvalue problem in an extended Hilbert space. The eigenvalue $\varepsilon(\mathbf{k})$ is the energy of Floquet-Bloch states in a system with periodic driving, dubbed Floquet-Bloch band structures. In fact, eigenvalues of $\varepsilon(\mathbf{k})$ and $\varepsilon(\mathbf{k})+n\hbar\omega$ represent the same Floquet state, and thus we could define the first Floquet Brillouin zone in ($-\hbar\omega/2, +\hbar\omega/2$). The states beyond the first Floquet Brillouin zone can be obtained through multiphoton absorption or emission from states inside the first Floquet Brillouin zone. Overall, the Floquet-Bloch bands form a series of photo-dressed replica bands, which can be deformed by light irradiation and induce hybridization with the folded bands at the first Floquet Brillouin zone boundary, resulting in the electronic and topological properties changed by coupling with light fields.

\subsection{Floquet Hamiltonian in tight-binding Wannier function}\label{subsec22}
To reveal the light-induced modification of crystalline materials, it is required to carry out first-principles calculations to obtain the basis of plane waves. By projecting plane waves of
Bloch states onto localized Wannier basis using
the WANNIER90 package~\cite{Mostofi2014}, we constructed real-space tight-binding Wannier Hamiltonian as
\begin{equation}
H^{W}=\sum_{m,n,\mathbf{R},\mathbf{R}'}t_{mn}(\mathbf{R}-\mathbf{R}')C_{m}^{\dag}(\mathbf{R})C_{n}(\mathbf{R}')+h.c.,
\end{equation}
where $\mathbf{R}$ and $\mathbf{R}'$ are lattice vectors, ($m$,$n$) is the index of Wannier orbitals, $t_{mn}(\mathbf{R}-\mathbf{R}')$ are the hopping integrals between Wannier orbital $m$ at site $\mathbf{R}$ and Wannier orbital $n$ at site $\mathbf{R}'$, and $C_m^{\dag}(\mathbf{R})$ or $C_m (\mathbf{R})$ creates or annihilates an electron of Wannier orbital $m$ on site $\mathbf{R}$.
When a time-periodic and space-homogeneous monochromatic light field is applied to a material, the time-dependent hopping is obtained by using the Peierls substitution \cite{PhysRevA.27.72,PhysRevLett.110.200403},
\begin{equation}
t_{mn}(\mathbf{R}-\mathbf{R}', \tau)=t_{mn}(\mathbf{R}-\mathbf{R}')e^{i\frac{e}{\hbar}\mathbf{A}(\tau)\cdot \mathbf{d}_{mn}},
\end{equation}
where $\mathbf{A}(\tau)$ is the time-dependent vector potential of an applied light-field, and $\mathbf{d}_{mn}$ is the related position vector between Wannier orbital $m$ at site $\mathbf{R}$ and Wannier orbital $n$ at site $\mathbf{R}'$. The corresponding light-driven operator is $C_{m}(\mathbf{R}, \tau)= \sum_{\alpha=-\infty}^{\infty} C_{\alpha m}(\mathbf{R})e^{i\alpha \omega \tau}$ with the Floquet operator $C_{\alpha m}(\mathbf{R})$ \cite{PhysRevLett.110.200403}.  In this case, the time-dependent $H^{W}(\tau)$ hosts both lattice and time translational symmetries, so we can map it onto a time-independent Hamiltonian according to the Floquet theory \cite{PhysRev.138.B979,PhysRevB.34.3625,PhysRevA.7.2203,Gesztesy1981JPA}. By carrying out a dual Fourier transformation, the static Floquet Hamiltonian can be expressed as
\begin{equation}\label{eq3}
H^F({\mathbf{k}}, \omega)=\sum_{m, n}\sum_{\alpha, \beta}[H_{mn}^{\alpha-\beta}({\mathbf{k}}, \omega)+\alpha\hbar \omega \delta_{mn}\delta_{\alpha \beta}]C_{\alpha m}^{\dag}(\mathbf{k})C_{\beta n}(\mathbf{k})+h.c.,
\end{equation}
where $\omega$ is the frequency of an incident light and thus $\hbar \omega$ represents the energy of photon, and the matrix $H_{mn}^{\alpha-\beta}({\mathbf{k}}, \omega)$ can be obtained by Wannier Hamiltonian as
\begin{equation}\label{eq4}
H_{mn}^{\alpha-\beta}({\mathbf{k}}, \omega)=\sum_{\mathbf{R}}\sum_{\mathbf{R}'}e^{i\mathbf{k}\cdot (\mathbf{R}-\mathbf{R}')} \bigg(\frac{1}{T}\int_{0}^{T}t_{mn}(\mathbf{R}-\mathbf{R}')e^{i\frac{e}{\hbar}\mathbf{A}(\tau)\cdot \mathbf{d}_{mn}}e^{i(\alpha-\beta)\omega \tau}d\tau \bigg).
\end{equation}
According to Eq. (\ref{eq4}), one can use first-principles calculations based on density functional theory to quantitatively simulate evolution of electronic properties in light-matter coupled materials. Beyond effective models, the Floquet-Wannier Hamiltonian obtained from density functional theory calculations with Floquet theorem can predict the specific light-driven electronic topology of crystalline materials in nonequilibrium.

\section{Light-driven topological semimetallic states from effective models}\label{sec3}
In this section, we review the recent developments in light-driven topological phases from different kinds of topological semimetallic phases based on effective models.

\subsection{Light-driven type-I, type-II, and hybrid NLSMs}\label{subsec31}
The general model Hamiltonian of the undriven NLSM with a single nodal ring has the form~\cite{Kim15PRL,Yu15PRL,WengH15PRB,Chan16PRB},
\begin{equation}
H_{0}=c_{i}k_{i}^{2}\sigma_{0}+\left( m_{0}-m_{i}k_{i}^{2}\right) \sigma
_{z}+v_{y}k_{y}\sigma_{y},  \label{H0}
\end{equation}
where $m_0, m_i (i=x,y,z)$ and $c_i$ are model parameters, $v_y$ is the
velocity along the $y$-axis, $k_i$ are the crystal momenta, $\sigma_i$ are
Pauli matrices and $\sigma_0$ is the identity matrix. Here, Einstein's
summation convention is used, where the repeated indices imply the summation.

\begin{figure}
\begin{center}
\includegraphics[width=0.8\textwidth]{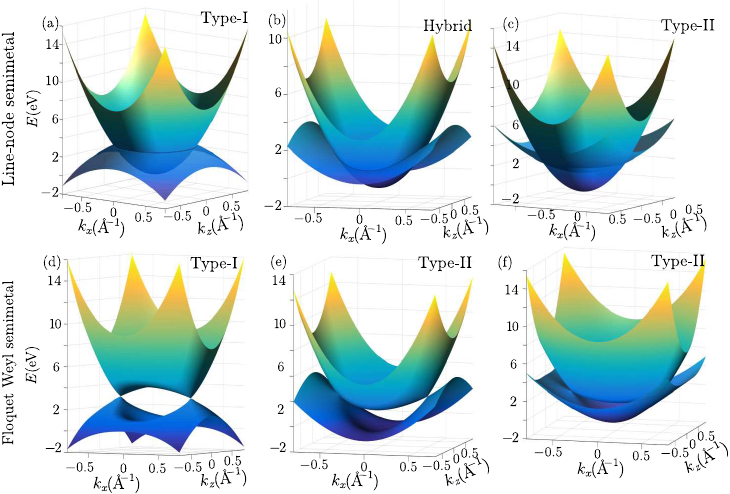}
\caption{Floquet WSMs phases can be induced in periodically driven NLSMs~\cite{Chen2018PRBLNSM}. (a-c) are the energy spectra of undriven type-I, hybrid and type-II NLSMs with (a) $\left\vert c_{x}/m_{x}\right\vert =\left\vert c_{z}/m_{z}\right\vert <1$, (b) $\left\vert c_{x}/m_{x}\right\vert >1$ and $\left\vert c_{z}/m_{z}\right\vert <1$, and (c) $\left\vert c_{x}/m_{x}\right\vert =\left\vert c_{z}/m_{z}\right\vert >1$, with $k_y=0$. (d-f) are the energy spectra of Floquet WSMs in driven type-I, type-II, and hybrid NLSMs by a light along the $x$-axis.}
\label{fig_Line_node}
\end{center}
\end{figure}

Depending on the parameters, the NLSM can be categorized into three types, as illustrated in Figs.~\ref{fig_Line_node}(a)-\ref{fig_Line_node}(c). In Fig.~\ref{fig_Line_node}(a), where the tilt is weak, the band touching exhibits a type-I nodal ring. Conversely,
Fig.~\ref{fig_Line_node}(c) shows that with strong tilt, both bands align in the same direction, forming a type-II nodal ring at their intersection.
Figure~\ref{fig_Line_node}(b) presents the band spectrum for a hybrid NLSM, revealing that the tilt ratio is smaller near the $k_z$-axis and larger near the $k_x$-axis.

To study the interaction of NLSMs with light,  a time-dependent
vector potential $\mathbf{A}\left( t\right) =\mathbf{A}\left( t+T\right) $ is considered,
which is a periodic function with a period of $T=2\pi/\omega$. Applying Floquet theory~\cite{PhysRev.138.B979,PhysRevB.34.3625,PhysRevA.7.2203,Gesztesy1981JPA}
in the high-frequency limit, the periodically driven system can be
described by a static effective Hamiltonian given by~\cite{Maricq1982PRB,Grozdanov1988PRA,Rahav2003PRA,Rahav2003PRL,Goldman,Eckardt2015NJP,Bukov2015AP}
\begin{equation}
H_\text{eff}=H_{0,0}+\frac{\left[ H_{0,-1},H_{0,1}\right]}{\hbar\omega}+O\left( A_{L}^{4}\right), 
\label{effective}
\end{equation}
where $\omega$ and $A_L$ describe the frequency and amplitude of light, $H_{m,n}=\frac{1}{T}\int_{0}^{T}H (t) e^{i(m-n)\omega t}dt $ is the discrete Fourier components of the Hamiltonian.
When the light propagates along the $x$-axis, $\mathbf{A}(t)$ is
given by $\mathbf{A}=A_{L}\left(0,\cos\omega t,\eta\sin\omega t\right)$,
where $\eta=\pm 1$ indicates the chiralities of CPL. From Eq. (\ref{effective}), the Floquet correction is
\begin{equation}
\Delta H^{x}=-\frac{A_{L}^{2}}{2}\left(m_{y}+m_{z}\right)\sigma_{z}-Lm_{z}k_{z}\sigma_{x},
\end{equation}
with $L=2\eta A_{L}^{2}v_{y}/(\hbar\omega)$. In the presence of the light, the coupling term gaps out the nodal ring except at two Weyl points $\pm\mathbf{k}_{0}=( \pm\sqrt{\tilde{m}_{0}/m_{z}},0,0)$ with $\tilde{m}_{0}=m_{0}-A_{L}^{2}\left( m_{y}+m_{z}\right)/2$.
The results indicate that a light traveling along the $x$-axis gaps out
the nodal ring, leaving a pair of Weyl nodes and causing the system enters into a WSM
phase~\cite{Chen2018PRBLNSM,Taguchi2016PRBa}. However, the type of the Weyl nodes is independent of the intensity
and the frequency of the incident light, suggesting that a type-II
Floquet WSM state arises by driving the type-II NLSM with a light along the $x$-axis.
The band spectrum of the driven type-I NLSM [Fig.~\ref{fig_Line_node}(d)] shows the type-I Weyl nodes. The bulk band spectra of the driven hybrid NLSM and the driven type-II NLSM are depicted in Figs.~\ref{fig_Line_node}(e) and \ref{fig_Line_node}(f), respectively. 

\begin{figure}[ptb]
\begin{center}
	\includegraphics[width=8cm]{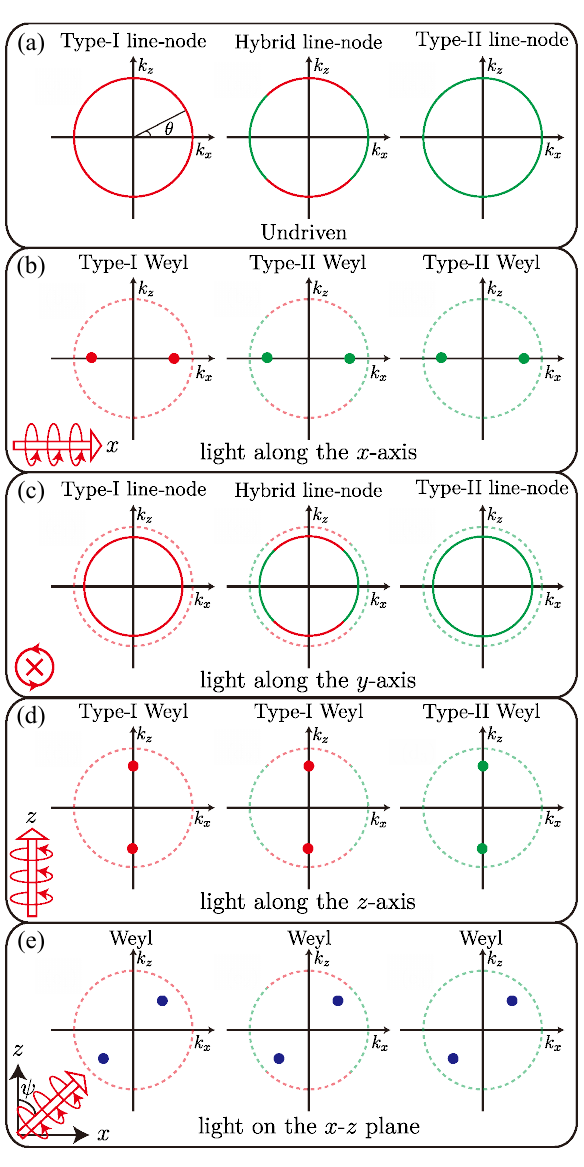}
	\caption{Schematics of driven type-I, hybrid, and type-II NLSM with different propagation directions of CPL~\cite{Chen2018PRBLNSM}. (a) Nodal rings of undriven NLSMs. (b) and (d) correspond to the case where the incident light travels along the $x$-axis and $z$-axis, respectively. The dots colored with red (or green) denote type-I (or II) Weyl nodes. Nodal rings are gapped out and Weyl nodes are created along the propagation direction of light. (c) The incident light along the $y$-axis only shifts nodal rings. (e) The light propagates on the $x$-$z$ plane with $\protect\psi$ defining the incident angle away from the $z$-axis. Nodal rings are gapped out and pairs of Weyl nodes appear along the propagation direction. The Weyl nodes are labeled with blue color as their types depend on model parameters, the incident angle, and the strength of light. The dashed lines in (b-e) correspond to nodal rings in undriven cases.}
	\label{section3fig1}
\end{center}
\end{figure}

Besides, it has been shown that the type of Floquet WSM phases depends on the orientation of the incident light~\cite{Chen2018PRBLNSM,Taguchi2016PRBa}. When the incident light propagates along the $x$-axis or along the $z$-axis, a type-II NLSM is converted into a type-II WSM 
, while for a driven hybrid NLSM, depending on the tilt direction, the photoinduced Floquet WSM could be of type-I [Fig.~\ref{section3fig1}(b)] or type-II [Fig.~\ref{section3fig1}(d)]. When the applied light propagates along the $y$-axis, only the positions of nodal rings change Fig.~\ref{section3fig1}(c) 
. Surprisingly, by rotating the incident light on the $x$-$z$ plane, both type-I and type-II WSMs can be realized by tuning the driving angle and amplitude Fig.~\ref{section3fig1}(e) 
. For the sake of comparison, the Figs.~\ref{section3fig1}(a)-\ref{section3fig1}(e) also give the Floquet states of driven type-I NLSMs by a CPL, which show different features from those of type-II and hybrid NLSMs. Furthermore, the anomalous Hall effects of these photoinduced Floquet WSM phases are also investigated by use of the Kubo formula~\cite{Chen2018PRBLNSM,Taguchi2016PRBa}.

\subsection{Light-induced higher-order WSM phases}\label{subsubsec32}
In this section, we show that CPL can also induce higher-order WSM phases that support both surface Fermi arcs and hinge Fermi arcs in a higher-order NLSM~\cite{XLDU2022} or a higher-order DSMs~\cite{ZMWangPRB}.

\begin{figure}[htpb!]
\begin{center}
	\includegraphics[width=0.8\textwidth]{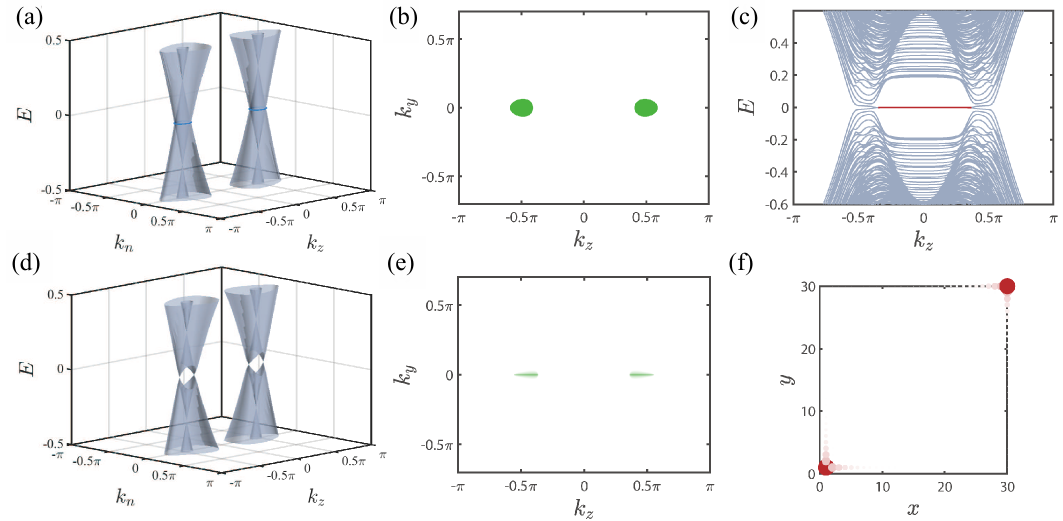}
	\caption{
		(a)-(c) Electronic structure of the undriven higher-order NLSM~\cite{XLDU2022}.  (d)-(f) Higher-order NLSM under CPL propagates along the $z$-axis~\cite{XLDU2022}. (a) The bulk energy spectrum as functions on the mirror plane. (b) The spectral density in the surface Brillouin zone. The drumhead surface states appear in the regions bounded by the two projected nodal rings. (c) The energy spectrum as a function of $k_z$ with the open boundary conditions along both the $x$ and $y$ directions. The red solid lines represent the hinge Fermi arc states. (d) The bulk energy spectrum as functions on the mirror plane under CPL driven.  (e) The location of surface Fermi arc states terminated at the projection of Weyl points on the surface Brillouin zone defined on the $k_y$-$k_z$ plane. The two line segments marked by green lines show the surface spectral density for $E = 0$. (f) The probability distribution of the hinge Fermi arc states at $k_z = 0$ under CPL driven. }
	\label{section3fig2}
\end{center}
\end{figure}

The Hamiltonian for the NLSM has the form~\cite{PhysRevLett.123.186401,PhysRevLett.125.126403,PhysRevLett.126.196402,PhysRevLett.121.106403,Lee2020},
\begin{align}
	H(\mathbf{k}) =& im(\Gamma_1 \Gamma_4 + \Gamma_2 \Gamma_4) + t\sin{k_x}\Gamma_1 + t\sin{k_y} \Gamma_2 \\ \nonumber
	& + [M-t(\cos{k_x}+\cos{k_y}+\cos{k_z})] \Gamma_3,
\end{align}
where the Dirac matrices are defined as $\Gamma_1=\sigma_0 \tau_3$, $\Gamma_2=\sigma_2 \tau_2$, $\Gamma_3=\sigma_0 \tau_1$, $\Gamma_4=\sigma_1 \tau_2$, where $\sigma_j$ and $\tau_j$ ($j=1,2,3$) are Pauli matrices labeling the sublattice and layer degrees of freedom, $\sigma_0$ and $\tau_0$ are identity matrices. $t$ is the amplitudes of hoppings, $M$ is the Dirac mass, and $m$ is the additional mass that generates higher-order topology. The system supports two bulk nodal rings, two-dimensional (2D) drumhead surface states, and one-dimensional hinge Fermi arc states. 
This model can describe the higher-order NLSM materials as $X$Te$_2$ ($X=$Mo, W)~\cite{PhysRevLett.123.186401} and 3D ABC stacked graphdiyne~\cite{PhysRevLett.121.106403,Lee2020,PhysRevLett.128.026405}.
Figure~\ref{section3fig2}(a) illustrates two mirror-protected bulk nodal rings on the $k_n$-$k_z$ mirror plane with $\mathbf{k_n}$ along the $k_x=-k_y$ axis in the first Brillouin zone. The drumhead surface states depicted in Fig.~\ref{section3fig2}(b) are the projection of bulk nodal rings on the $k_y$-$k_z$ plane. Figure~\ref{section3fig2}(c) demonstrates the hinge Fermi arc states, which are located on two mirror-symmetric off-diagonal hinges.

\begin{figure}[htpb!]
\begin{center}
	\includegraphics[width=0.8\textwidth]{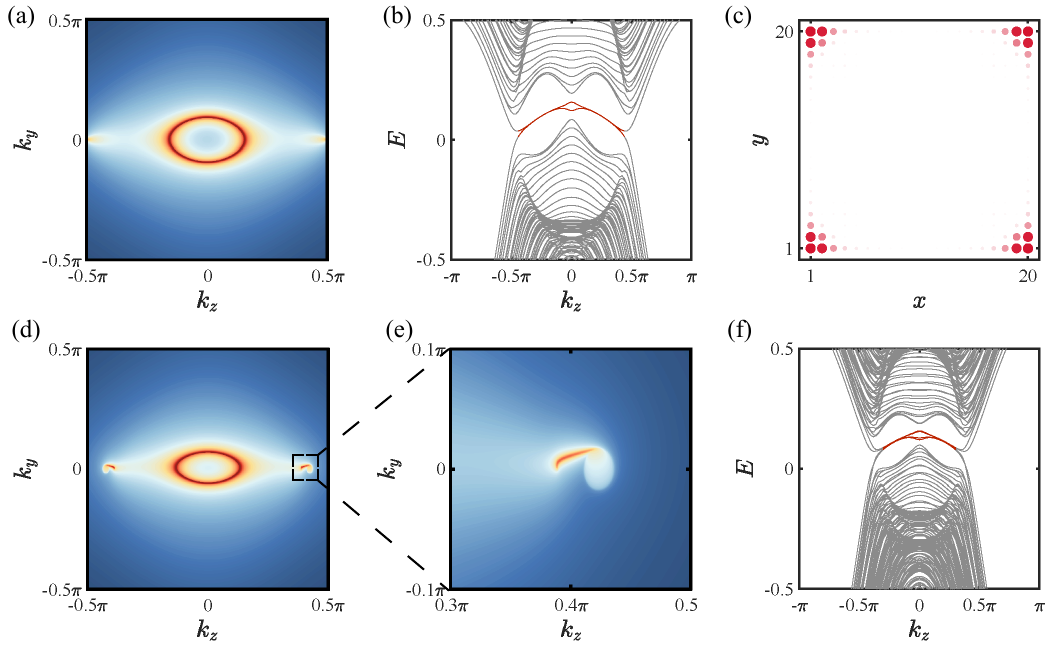}
	\caption{
		(a)-(c) Electronic structure of the undriven higher-order DSM~\cite{ZMWangPRB}. (d)-(f) Higher-order DSM under CPL propagates along the $z$-axis~\cite{ZMWangPRB}. (a) The surface spectral function on the $k_y$-$k_z$ plane with the semi-infinite boundary along the $x$ direction when $E = 0$. (b) The energy spectrum vs $k_z$ with the open boundary condition along both the $x$ and $y$ directions. The solid red lines represent the topologically protected hinge Fermi arc states. (c) The local density of states (LDOS) of the hinge Fermi arc states at $k_z = 0.1\pi$. (d) The surface spectral function on the $k_y$-$k_z$ plane under CPL driven. The red solid lines around the original Dirac points mark the surface Fermi arcs, which connect the same pair of Weyl points. The red closed circle shows the surface Fermi rings. (e) The zoom-in view of the surface Fermi arc within the black dashed square in (d).  (f) The hinge states are marked by the solid red lines under CPL driven. }
	\label{section3fig3}
\end{center}
\end{figure}

When the CPL propagates along the $z$-axis, $\mathbf{A}(t)$ is given by $\mathbf{A}=A_L(\eta \sin{\omega t}, \cos{\omega t},0)$. The effective Hamiltonian of the driven higher-order NLSM can be found in Ref. \cite{XLDU2022}.
The light irradiation breaks both the $\mathcal{T}$-symmetry and chiral symmetry, gapping out the nodal rings and leaving a pair of Weyl nodes, as shown in Fig.~\ref{section3fig2}(d). Figure~\ref{section3fig2}(e) demonstrates that the surface drumhead states are replaced by surface Fermi arc. However, the mirror symmetry and the product of time-reversal and chiral symmetries are still preserved, which protect the higher-order hinge Fermi arcs shown in Fig.~\ref{section3fig2}(f). The above results indicate that the driven system turns into a higher-order WSM, which supports both first-order surface Fermi arc and second-order hinge Fermi arc states.
Moreover, when light propagates along the other axis, CPL can always drive a higher-order NLSM to a higher-order WSM. More importantly, it is found that the propagation axis of CPL can control the location of the Weyl nodes.

A similar conclusion is found in the CPL-driven higher-order DSM~\cite{ZMWangPRB}. The model describing an undriven higher-order DSM model is given by~\cite{Mao2018PRB,wieder2020strong,yang2021classification,Tyner2021quantized,Nie2022HODSM,zeng2022topological,HOTDSM-EXP,WANG2022788,Wang2020higher-order,Ghorashi2020higher-order,Roy2019antiunitary}:
\begin{align}
	H(\mathbf{k}) =& \epsilon_0(\mathbf{k}) + \lambda \sin{k_x} \Gamma_1^{\prime} + \lambda \sin{k_y} \Gamma_2^{\prime} \\ \nonumber
	& + M(\mathbf{k})\Gamma_3^{\prime} + G(\mathbf{k})\Gamma_4^{\prime},
\end{align}
where $\epsilon_0(\mathbf{k})=t_1 (\cos{k_z}-\cos{K_z^0})+t_2(\cos{k_x}+\cos{k_y}-2)$, $M(\mathbf{k})=t_z (\cos{k_z}-\cos{K_z^0})+t(\cos{k_x}+\cos{k_y}-2)$, $\Gamma_1^{\prime}=s_3 \rho_1$, $\Gamma_2^{\prime}=s_0 \rho_2$, $\Gamma_3^{\prime}=s_0 \rho_3$, $\Gamma_4^{\prime}=s_1 \rho_1$, $\Gamma_5^{\prime}=s_2 \rho_1$, $s_j$ and $\rho_j$ ($j=1,2,3$) are Pauli matrices denoting the spin and orbital degrees of freedom, and $s_0$ and $\rho_0$ are identity matrices. This model can describe the higher-order DSM materials including but not limited to Cd$_3$As$_2$ and KMgBi~\cite{wieder2020strong}.  Here, $t$, $\lambda$, and $t_{1,2,z}$  are the amplitudes of hoppings. The two Dirac cones are located at $\mathbf{k}=(0,0,\pm K_z^0)$, as shown at the boundary of Fig.~\ref{section3fig3}(a). 
$G(\mathbf{k})=g(\cos{k_x}-\cos{k_y})\sin{k_z}$ represents the higher-order topological term, which gives rise to second-order hinge Fermi arc states as displayed in Fig.~\ref{section3fig3}(b)-\ref{section3fig3}(c).
This system has different surface states, as demonstrated in Fig.~\ref{section3fig3}(a). The closed Fermi ring, instead of helical Fermi arc states, emerges in the surface Brillouin zone~\cite{SFAofDSM} when the Fermi energy cuts through the surface Dirac cone.


The CPL also drives the higher-order DSM into a Floquet WSM, separating each Dirac point into a pair of Weyl points. These Weyl points host surface Fermi arc states that connect the projections of each pair of Weyl nodes, as depicted in Fig.~\ref{section3fig3}(d)-\ref{section3fig3}(e). Moreover, Fig.~\ref{section3fig3}(f) shows that the Floquet WSM also hosts hinge Fermi arc states, terminated by the projection of two adjacent Weyl nodes from two different pairs.
Moreover, CPL can drive the tilted Weyl cones in Floquet WSMs, when the axis of light propagation is changed. On the other hand, the higher-order term also can be written as $G(\mathbf{k})=g(\cos{k_x}-\cos{k_y})$; this term also breaks the four-fold rotational symmetry, but preserves the effective parity-time reversal ($\mathcal{PT}$) symmetry. In this case, the CPL plays a similar role in inducing the topological phase in the higher-order DSM.

The above content focuses on the higher-order topological semimetals with the four-fold rotational symmetry. In the higher-order DSM with the six-fold rotational symmetry, CPL irradiation can also produce Floquet WSM that supports both first-order surface Fermi arc and higher-order hinge Fermi arc states, and the location of Weyl nodes depends on the propagation direction of angle of incidence of CPL~\cite{Xu2024CPB}. Furthermore, CPL can also control the degree of tilt of the
resulting Weyl cones by adjusting the incident direction of the CPL, enabling the realization of different types of WSMs.

\section{Topological states induced by the interplay of periodic driving and disorder}\label{sec4}
The investigation of the condensed matter systems in the presence of disorder is a longstanding research area of fundamental importance. For instance, the disorder is crucial for the observation of integer quantized Hall plateau~\cite{IQHE1980,Halperin1982,QHEbook}. Over the past two decades, significant progress has been made in the study of disorder effects in topological systems ~\cite{TAI-1,TAI-2,3DTAI,expTAI,PTAI,Titum1,Titum2,Liuprl2020TAI}. Surprisingly, despite the presence of disorder-induced localization effects, the disorder may also endow a system with non-trivial topological states i.e., the TAI~\cite{TAI-1}. 
The interplay of disorder and periodic driving can give rise to rich topological phenomena~\cite{Titum1,Titum2,wauters2019prl,PTAI}. In this section, we review the recent works about topological states induced by the interplay of periodic driving and disorder. 

\subsection{Light-induced QAH effects in disordered systems}\label{subsec41}
\begin{figure}[htb]
    \centering
    \includegraphics[width=1\textwidth]{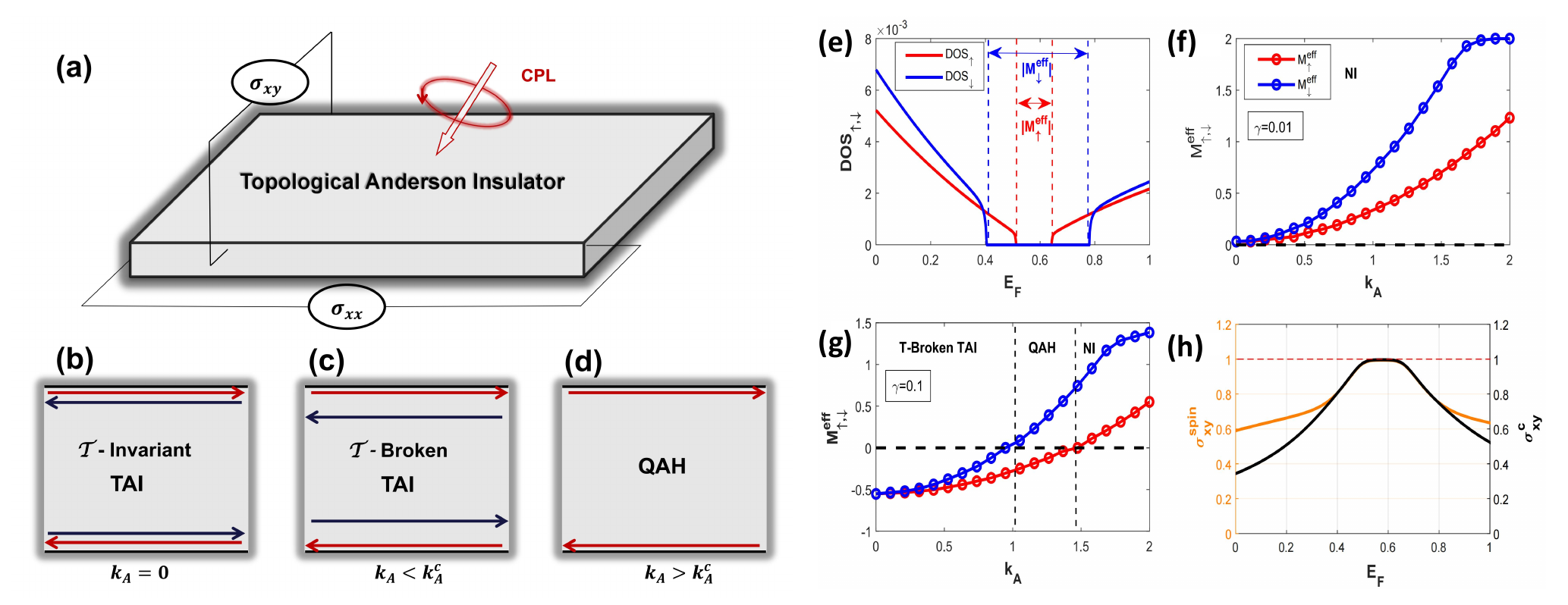}%
    \caption{
    Light-induced QAH effects in disorder systems~\cite{PhysRevB.105.035103}. (a)-(d) The schematic diagram of QAH state in CPL-driven TAI.
    (e) The density of states of BHZ model in the presence of CPL and disorder, the strength of disorder is $\gamma = 0.1$ and the intensity of CPL is $k_A = 1.25$. (f)-(g) The effective mass as a function of light intensity $k_A$. The strength of disorder is (f) $\gamma = 0.01$, (g) $\gamma = 0.1$, (h) The spin Hall conductivity $\sigma^{spin}_{xy}$  (orange solid line) and the charge Hall conductivity $\sigma^{c}_{xy}$ (black solid line). The charge conductivity is in the unit of ${e^2}/{h}$ and the spin conductivity is in the unit of ${e}/{4\pi}$.}
    \label{fig1}
\end{figure}

In the original version of TAI, $\mathcal{T}$-symmetry is preserved and the phenomenology is similar to the quantum spin Hall effect. Under the breaking of $\mathcal{T}$-symmetry, the QAH state can be realized~\cite{Oka1,Inoue,Oka2,Ezawa,Mimp}. A natural analogy is as follows: Can we induce the QAH effect in disordered systems? In the following, we briefly review that the QAH state arises from the interplay of light and nonmagnetic disorder. Here, a strategy was proposed to induce QAH in disordered systems~\cite{PhysRevB.105.035103}. The idea is to explore systems in which bulk topology is driven by the nonmagnetic disorder, and the spin degeneracy is lifted under irradiation of light fields, as shown in Fig. \ref{fig1}. As the light intensity increases beyond a critical value $k_{A}^{c}$, the energy gap of one spin sector first closes and then reopens, while the other spin sector still possesses the nontrivial topology. In this case, the system evolves into the QAH phase with one gapless chiral edge channel [Figs. \ref{fig1}(b)-(d)].

The four-band Bernevig-Hughes-Zhang (BHZ) effective Hamiltonian~\cite{RevModPhys.83.1057,BHZ-2,BHZ-3} can be used to reveal the light-modulated topological phases mentioned above.
\begin{equation}
\begin{aligned}
h_{s}(\mathbf{k}) = d_0(\mathbf{k})\tau_0+\mathbf{d}_{s}(\mathbf{k}) \cdot \boldsymbol{\sigma},
\end{aligned}
\label{eq:BHZ}
\end{equation}
where the index $s=\pm$ denotes spin, or the inequivalent valleys $\pm K$ due to spin-valley locking in mori\'{e} superlattices. For the low-energy effective minimal model, we have $d_0(\mathbf{k}) = -Dk^2$ and $\mathbf{d}_{s}(\mathbf{k}) = (sAk_x,Ak_y,M-Bk^2)$, where $M$ is the Dirac mass term depicting the band inversion, the other parameters $A, B, D$ can be obtained from experiments. The sign of Dirac mass term is crucial to characterize the topological phase transition from normal insulator ($M>0$) to topological insulator ($M<0$).

In the presence of CPL and disorder, the Hamiltonian becomes time-dependent and the translational symmetry is broken by lattice disorder. However, a simple picture can still be obtained by the following approximations. In off-resonant regime (i.e., high frequency), the time-independent effective Hamiltonian can be approximately obtained through the Magnus expansion~\cite{Goldman}. On the other hand, the effects of disorder can be taken into account by effective medium theory, such as the self-consistent Born approximation (SCBA)~\cite{TAI-2,bruusMB}. After considering these approximations, the renormalized Dirac mass can be expressed as:
\begin{equation}
M^{\rm{eff}}_s = M - \Delta^{cpl}_s(k_A) - \Sigma^{dis}_s(\gamma),
\label{eq:DM}
\end{equation}
where the second term on the right hand side of Eq.~(\ref{eq:DM}) $\Delta^{cpl}_s(k_A)$ is a function of light intensity $k_A$ induced by CPL. The third term $\Sigma^{dis}_s(\gamma)$ is disorder induced self-energy which can be calculated by SCBA for a given disorder strength $\gamma$. It is noted that both corrections $\Delta^{cpl}_s, \Sigma^{dis}_s$ are spin-dependent, leading to two spin sectors with different responses to CPL.
The effective mass $M^{\rm{eff}}_s$ can be extracted from the width of the quasi-band band gap, as shown in Fig. \ref{fig1}(e). If the disorder is too weak to induce the TAI phase, the increasing of the light-intensity $k_A$ cannot induce any topological phase transitions [Fig. \ref{fig1}(f)]. Therefore, one need a moderate strength of disorder to drive the system into TAI. Specifically, as shown in Fig. \ref{fig1}(g), the effective mass is negative in the absence CPL, $M^{\rm{eff}}_s(k_A=0)<0$. Once irradiation of CPL is applied, two spin sectors exhibit different responses, and the system shows $\mathcal{T}$-symmetry broken TAI when $M^{\rm{eff}}_{\uparrow,\downarrow}<0$. As the light intensity exceeds a critical value $k_{A}^{c}$, the energy gap of one spin sector first closes and then reopens $M^{\rm{eff}}_{\downarrow}>0$, while the other spin sector still possesses the nontrivial topology $M^{\rm{eff}}_{\uparrow}<0$ [Fig. \ref{fig1} (g)]. In this case, the system evolves into the QAH phase. The spin-polarized topological phases can also be characterized by the spin Hall conductivity $\sigma^{spin}_{xy}$ and charge Hall conductivity $\sigma^{c}_{xy}$ [Fig. \ref{fig1}(h)]. These results conceptually demonstrate the possibility of realizing QAH effect in TAI using CPL.

\subsection{Quadrupole topological insulator with periodic driving and disorder}
Recently, the concept of topological phase of matter has been extended to higher-order topological phases~\cite{BBH,BBH2017PRB,Langbehn2017PRL,Song2017PRL,Schindler2018SA,xie2021higher}. 
Among various higher-order topological phases, the quadrupole topological insulator (QTI)~\cite{BBH} associated with a quantized quadrupole moment is of particular interest, which accommodates topologically protected corner states. It has been believed that spatial symmetries (such as mirror symmetries and/or four-fold rotation symmetry) and internal symmetries (such as chiral symmetry, $\mathcal{T}$-symmetry and particle-hole symmetry) are crucial ingredients to design and realize QTIs. The presence of disorder explicitly breaks crystal symmetries. Therefore, the new phenomena induced by disorder in the higher-order topological states have attracted wide attention ~\cite{li2020prl,zhang2021prl,yang2021prb}.  
In the following, we review that an exotic QTI created by the intertwined periodic driving and disorder emerges from a topologically trivial band structure. This intriguing QTI possesses a quantized quadrupole moment only protected by particle-hole symmetry.

\begin{figure}[htb]
\begin{center}
	\includegraphics[width=0.7\textwidth]{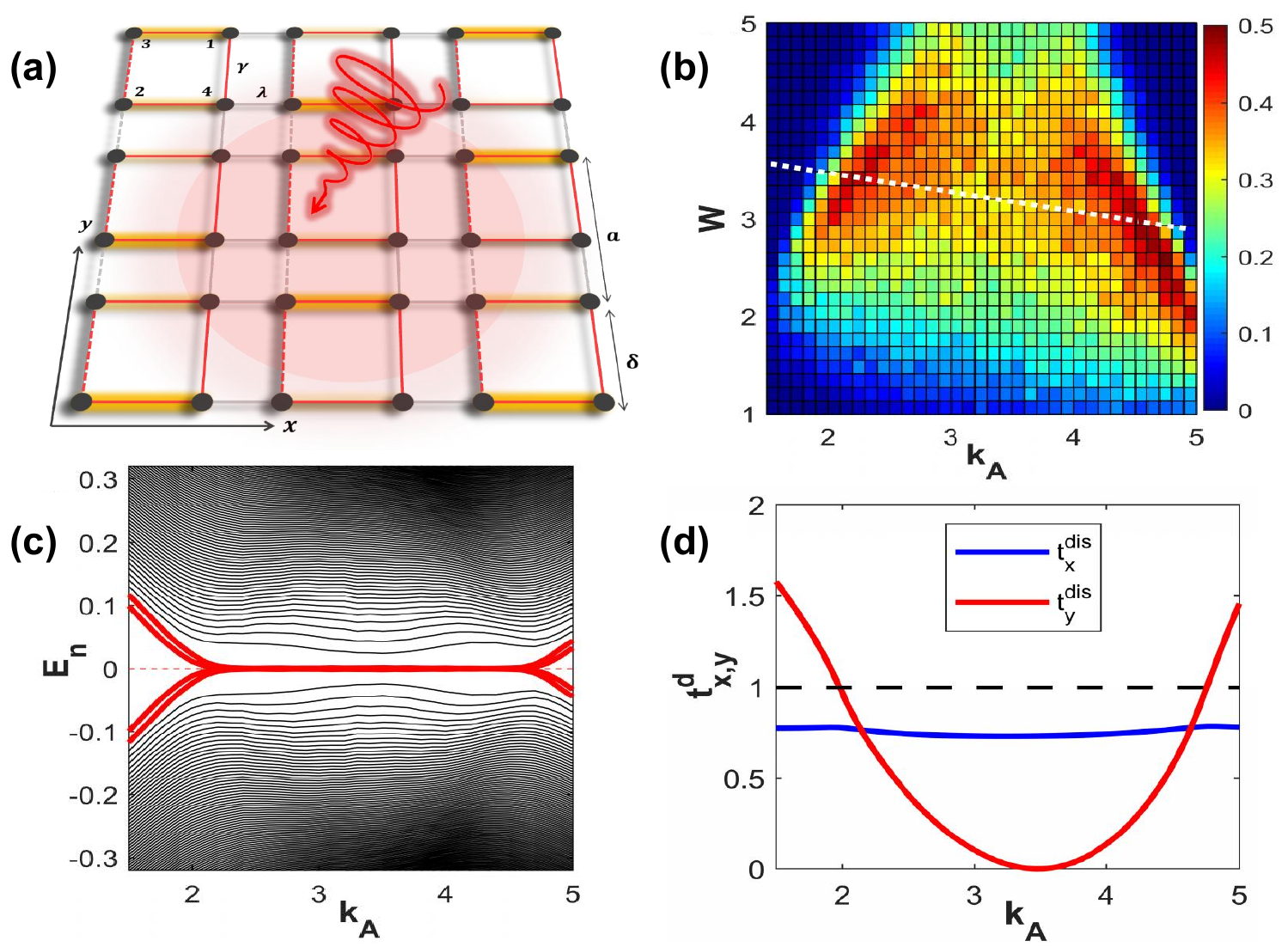}
	\caption{QTI with periodic driving and disorder \cite{PhysRevB.105.L201114}. (a) The schematic illustration of the BBH lattice model with the lattice constant $a$ under periodic driving. The distance between the nearest neighbor in a unit cell is denoted by $\delta$. (b) The phase diagram on the $W-k_A$ parameter plane depicted by the average value of quadrupole moment $Q_{xy}$, where we denote the strength of disorder and the strength of light field by $W$ and $k_A$ respectively. (c) The energy spectrum $E_n$ with open boundaries. (d) The effective hopping amplitudes $t^{\rm{d}}_{x,y}$ as a function of $k_A$.}
	\label{Fig:FHOTAI}
\end{center}
\end{figure}

Starting with the Benalcazar-Bernevig-Hughes (BBH) model, a paradigmatic model of QTIs, the Hamiltonian can be written as
\begin{equation}
\begin{aligned}
&H_q({\mathbf{k}}) =  \lambda\sin(k_y)\tau_2\sigma_1 +\lambda\sin(k_x)\tau_2\sigma_3 \\
&[\gamma + \lambda\cos(k_x)]\tau_1\sigma_0 + [\gamma + \lambda\cos(k_y)]\tau_2\sigma_2.
\end{aligned}
\label{eq:BBH}
\end{equation}
The schematic illustration of the lattice structure is shown in Fig. \ref{Fig:FHOTAI} (a).
The topological phase transition is determined by the ratio of $\gamma / \lambda$. In the static and clean limit, the BBH model Eq. (\ref{eq:BBH}) describes a trivial insulator when $\gamma / \lambda>1$ or a QTI when $\gamma / \lambda<1$ . The phase boundary is at $\gamma / \lambda = 1$. Under the irradiation of CPL, electronic structures of the system are effectively modified by the virtual photon absorption processes, which can be expressed as an effective Floquet-Bloch Hamiltonian:
\begin{equation}
\begin{aligned}
H^{\rm{eff}}({\mathbf{k}}) &= H_q({\mathbf{k}}) + H'({\mathbf{k}}),
\end{aligned}
\label{eq:HFK}
\end{equation}
where the second term $H'({\mathbf{k}})$  induced by CPL gives an important modification to the original BBH model~\cite{PhysRevB.105.L201114}. It is noted that chiral symmetry and $\mathcal{T}$-symmetry preserved by the static BBH Hamiltonian $H_q({\mathbf{k}})$ are both broken under the CPL. Nevertheless, the combination of these two symmetries (i.e., particle-hole symmetry) is preserved.
Particle-hole symmetry is critical to the quantization of quadrupole moment in the even presence of both disorder and periodic-driving.
The quadrupole moment defined in real space $Q_{xy}$~\cite{li2020prl,yang2021prb,QMR1,QMR2,Roy2019antiunitary} can characterize the QTI phase in disordered systems. When the periodic driving and disorder are simultaneously present, all crystalline symmetries and chiral symmetry are destroyed. However, the preserved particle-hole symmetry can protect the quantization of quadrupole moment~\cite{li2020prl}. The established topological invariant $Q_{xy}$ allows one to investigate topological phase transitions in the presence of periodic driving and disorder.

Figure \ref{Fig:FHOTAI}(b) depicts the phase diagram of the quadrupole moment $Q_{xy}$ on the $W-k_A$ parameter plane. Significantly, there is an explicit area of the QTI phase, which is created by the joint effort of periodic driving and disorder. This QTI phase can neither be created by the driving field without disorder, nor disorder in the absence of driving field. For example, in Fig. \ref{Fig:FHOTAI}(b), the QTI phase cannot be induced by tuning the field strength $k_A$ in the weak disorder regime ($W\sim1$ ) or by increasing the disorder strength in the presence of weak driving field ($k_A \sim1$). 
The energy eigenstates by directly diagonalizing the tight-binding Hamiltonian with open boundaries can fully illustrate the characteristic features of the emergent QTI phase. As evident in Fig.~\ref{Fig:FHOTAI}(c), four in-gap modes at $E=0$ emerge as a function of $k_A$, implying that the presence of topologically nontrivial QTI phase. The zero-energy modes in bulk gap correspond to corner states. Furthermore, the topological phase transitions induced by the interplay of periodic driving and disorder can be understood by a simple picture based on the effective medium theory. The hopping amplitudes renormalized by periodic driving and disorder ($\gamma_{x,y} \rightarrow t^{\rm{d}}_{x,y} $) are displayed in Fig.~\ref{Fig:FHOTAI}(d). When $t^{\rm{d}}_{x}<1$ and $t^{\rm{d}}_{y }<1$ simultaneously, the disorder-induced Floquet QTI phase is created. This picture agrees with the numerical computations. The intriguing QTI phase, protected only by particle-hole symmetry, which necessitates the simultaneous presence of disorder and periodic driving further enriches the symmetry-protected mechanism of higher-order topology.

\section{Floquet engineering of topological states in realistic materials}\label{sec5}
In this section, we review the progress of light-driven QAH states and controllable Weyl fermions proposed using first-principles calculations combined with Floquet theory.


\subsection{The realization of light-driven QAH and VQAH states}\label{subsec51}

\begin{figure*}
\centering
\includegraphics[width=0.9\textwidth]{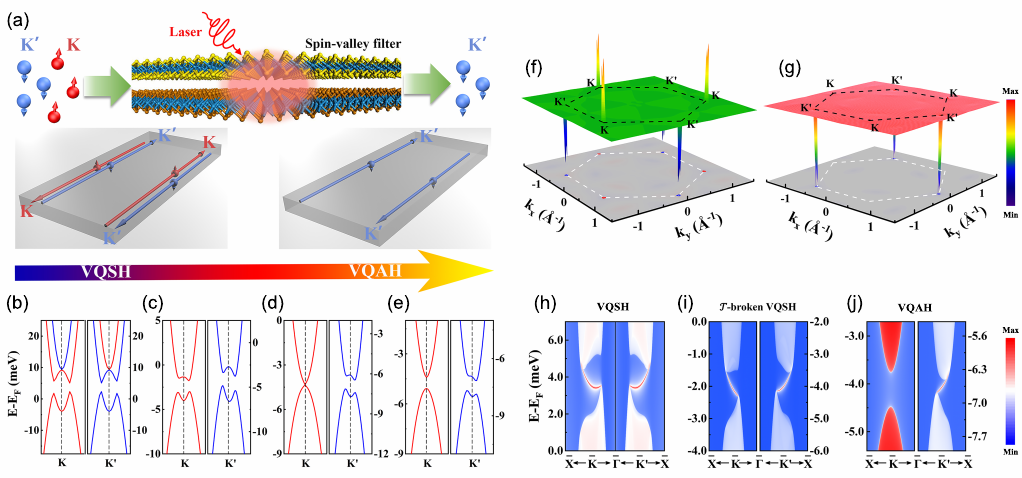}
\caption{Floquet VQAH state in nonmagnetic heterobilayers \cite{PhysRevB.105.L081115}. (a) Schematic illustration of an optically switchable topological spin-valley filter of TMD heterobilayers. (b)-(e) The evolution of spin-resolved band structures of WS$_2$/WTe$_2$ heterobilayer around the $K$ and $K'$ valleys with a light intensity $eA/\hbar$ of 0, 0.045, 0.051, and 0.053 {\AA}$^{-1}$. The red and blue colors indicate the $z$-component of spin-up and spin-down states, respectively. The distribution of the Berry curvature $\Omega_z(\mathbf{k})$ for (f) $\mathcal{T}$-broken (or invariant) VQSH state and (g) VQAH state in the $k_x-k_y$ plane, respectively. The hexagonal Brilllouin zone is marked by dashed lines. The calculated LDOS projected on a semi-infinite ribbon of zigzag edge with a light intensity $eA/\hbar$ of (h) 0, (i) 0.045, and (j) 0.053 {\AA}$^{-1}$, which respectively corresponds to $\mathcal{T}$-invariant VQSH state,  $\mathcal{T}$-broken VQSH state, and VQAH state.}\label{fig-TMD}
\end{figure*}

The light-driven QAH states can be realized in 2D nonmagnetic $MX_2$/WTe$_2$ ($M$=Mo, W; $X$= S, Se) transitional metal dichalcogenides (TMDs) heterobilayers under the irradiation of CPL. Considering the locking of spin to valley in TMDs, the valley-polarized quantum anomalous Hall (VQAH) state with one spin- and valley-resolved chiral edge channel in TMDs heterobilayers under light irradiation behaves as a perfect topological spin-valley filter [Fig. \ref{fig-TMD}(a)]. Figures. \ref{fig-TMD}(b)-\ref{fig-TMD}(e) illustrate the evolution of spin-resolved band structures around the $K$ and $K'$ valleys under CPL irradiation. We can find that the bands of spin-up states around the $K$ valley have been more drastically modified than those of spin-down states around the $K'$ valley. More importantly, with increasing light intensity, the band gap of spin-up states first closes and then reopens; that is, only spin-down states preserve the inverted band topology, resulting in a valley quantum spin Hall (VQSH) to VQAH topological phase transition. 
By integrating the Berry curvature as shown in Figs. \ref{fig-TMD}(f) and \ref{fig-TMD}(g), one can obtain $\mathcal{C}_K=1$ and $\mathcal{C}_{K'}=-1$, and the valley Chern number is $\mathcal{C}_v= 2$. For the VQAH state, the Berry curvature $\Omega_z$ only distributes and diverges near $K'$ valleys, giving $\mathcal{C}_K=0$ and $\mathcal{C}_{K'}=-1$, and then the Chern number is $\mathcal{C}=-1$. The topological phase with a specific topological invariant gives rise to uniquely nontrivial edge states. 
Without light irradiation, the $\mathcal{T}$-invariant VQSH state shows that two opposite chiral edge states with Kramers degeneracy are visible at the $K$ and $K'$ valleys [Fig. \ref{fig-TMD}(h)]. With increasing light intensity to 0.045 {\AA}$^{-1}$, the $\mathcal{T}$-broken VQSH state removes the Kramers degeneracy and exhibits different inverted band gaps at the $K$ and $K'$ valleys, but the chiral edge states are also visibly present [Fig. \ref{fig-TMD}(i)], confirming its nontrivial feature. As shown in Fig.~\ref{fig-TMD}(j), the VQAH state possesses one chiral edge state connecting the valence and conduction bands around the $K'$ valley, while bands around the $K$ valley exhibit the topologically trivial NI state. Furthermore, depending on light helicities, this CPL can selectively switch the states between two valleys and spin, providing a reliable scheme to realize an optically switchable topological spin-valley filter~\cite{PhysRevB.105.L081115}.

\begin{figure*}
\centering
\includegraphics[width=0.9\textwidth]{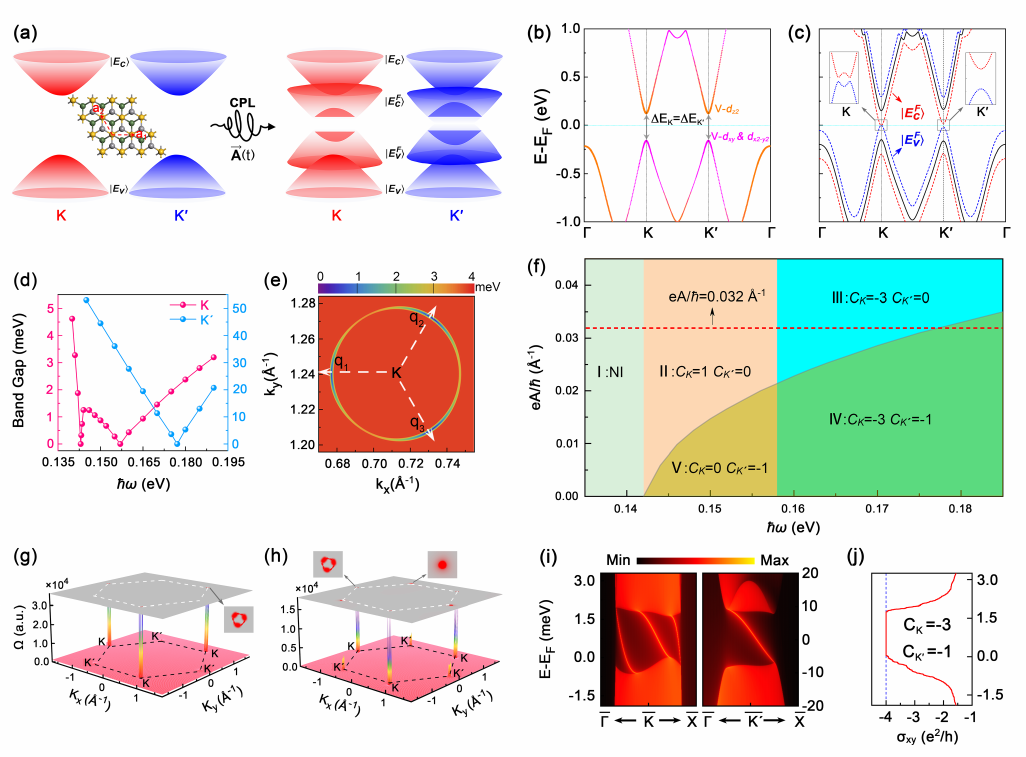}
\caption{Floquet engineering of nonequilibrium VQAH effect with tunable Chern number \cite{nanolett.2c04651}. (a) Band diagram illustrating valley-dependent band inversion via hybridization of Floquet-Bloch bands (i.e., $E_c^{F}$ and $E_v^{F}$) originated from irradiation of CPL. (b) The orbital-resolved electronic band structure of monolayer VSi$_2$N$_4$ including spin-orbital coupling (SOC). The components of V $d_{xy}+d_{x^{2}-y^{2}}$ and $d_{z^2}$ orbitals are proportional to the width of the magenta and orange lines, respectively. (c) The photon-dressed band structures of VSi$_2$N$_4$ subject to left-handed CPL with a certain light intensity and frequency. The black solid lines represent the equilibrium bands. The blue and red dashed lines represent Floquet-Bloch bands created by absorption and emission of photons, respectively. The insets indicate that two Floquet-Bloch bands invert at the $K$ point and preserve the trivial band gap at the $K'$ point. (d) The variation of band gaps around the $K$ and $K'$ valleys as a function of photon energy. (e) The energy difference map between Floquet-Bloch bands $E_c^{F}$ and $E_v^{F}$ in the vicinity of $K$ valley with a critical photon energy $\hbar\omega=0.157$ eV. (f) The phase diagram as functions of light intensity and frequency. Five distinct topological phases are shown as different colors. (g)-(h) The distribution of the Berry curvature under irradiation of left-handed CPL, which respectively correspond to the nontrivial regimes III and IV in panel (f). (i) The calculated semi-infinite LDOS, and (j) anomalous Hall conductance under irradiation of left-handed CPL with photon energy $\hbar\omega = 0.185$ eV and light intensity $eA/\hbar$ = 0.025 {\AA}$^{-1}$.}\label{fig-VSiN}
\end{figure*}

The Floquet QAH states can be further obtained from topologically trivial semiconductors. As illustrated in Fig. \ref{fig-VSiN}(a), under irradiation of CPL, periodic driving gives rise to Floquet-Bloch bands, and then two certain bands [i.e., labeled as $E_c^{F}$ and $E_v^{F}$ in Fig. \ref{fig-VSiN}(a)] move close to the Fermi level via the optical Stark effect \cite{2015Valley,PhysRevB.97.045307,nanolett.6b04419,science.aal2241}. 
This proposal can be implemented in the 2D semiconductors $M$Si$_2$$Z_4$ ($M$ = Mo, W, V; $Z$ = N, P, As) family materials. 
This family of materials, including magnetic and nonmagnetic members, host excellent stability, and especially MoSi$_2$N$_4$ and WSi$_2$N$_4$ were successfully synthesized in experiments \cite{science.abb7023}. As a representative example, the band structures of VSi$_2$N$_4$, with the valence band and conduction band respectively contributed by $d_{xy}$ $\&$ $d_{x^{2}-y^{2}}$ and $d_{z^2}$ orbitals of V atoms, depict a trivial semiconducting feature with the valley degeneracy in Fig. \ref{fig-VSiN}(b). Under light irradiation, in addition to the equilibrium bands (black solid lines), one can find that the Floquet-Bloch bands in Fig. \ref{fig-VSiN}(c), that are created by absorption (red dashed lines) or emission (blue dashed lines) of a photon, are present. 
Interestingly, the band gaps near the $K$ and $K'$ points indicate that band gap at the $K$ point closes and reopens twice [Fig. \ref{fig-VSiN}(d)]. Due to the threefold rotational ($C_3$) symmetry, a triangular distortion of the Fermi surface around the $K$ points would be present [Fig. \ref{fig-VSiN}(e)], which is known as trigonal warping \cite{PhysRevLett.98.176806,PhysRevB.82.113405,PhysRevB.95.045424,PhysRevB.101.161103,PhysRevB.104.195427}. 
The presence of trigonal warping would like to strongly enrich topological phases. As shown in Fig. \ref{fig-VSiN}(f), the phase diagram characterized by $\mathcal{C}_{K (K')}$ as functions of $\hbar \omega$ and $eA/\hbar$ indicates that there are five distinct topological phases, such as regime I: $\mathcal{C}_K = 0$ and $\mathcal{C}_{K'}=0$, regime II: $\mathcal{C}_K = +1$ and $\mathcal{C}_{K'}=0$, regime III: $\mathcal{C}_K = -3$ and $\mathcal{C}_{K'}=0$, regime IV: $\mathcal{C}_K = -3$ and $\mathcal{C}_{K'}=-1$, and regime V: $\mathcal{C}_K = 0$ and $\mathcal{C}_{K'}=-1$. Except the topologically trivial regime I, other four regimes are all related to the topologically nontrivial VQAH state. The Berry curvature distributions for regimes III and IV are plotted in Figs. \ref{fig-VSiN}(g) and \ref{fig-VSiN}(h). The nonzero Berry curvature $\Omega_z(\mathbf{k})$ diverges near the $K$ and/or $K'$ points. In particular, the $\Omega_z(\mathbf{k})$ near the $K$ point exhibits the $C_3$-symmetry as shown in the insets of Figs. \ref{fig-VSiN}(g) and \ref{fig-VSiN}(h), further confirming nontrivial band topology associated with trigonal warping. The Floquet VQAH states with specific first Chern number $\mathcal{C}$ and valley-resolved Chern number $\mathcal{C}_{K (K')}$ correspond to the valley-dependent chiral edge channels [Fig. \ref{fig-VSiN}(i)] and quantized Hall conductance $\sigma_{xy}$ [Fig. \ref{fig-VSiN}(j)], characterizing the global band topology of VQAH states.

\begin{figure*}
\centering
\includegraphics[width=0.9\textwidth]{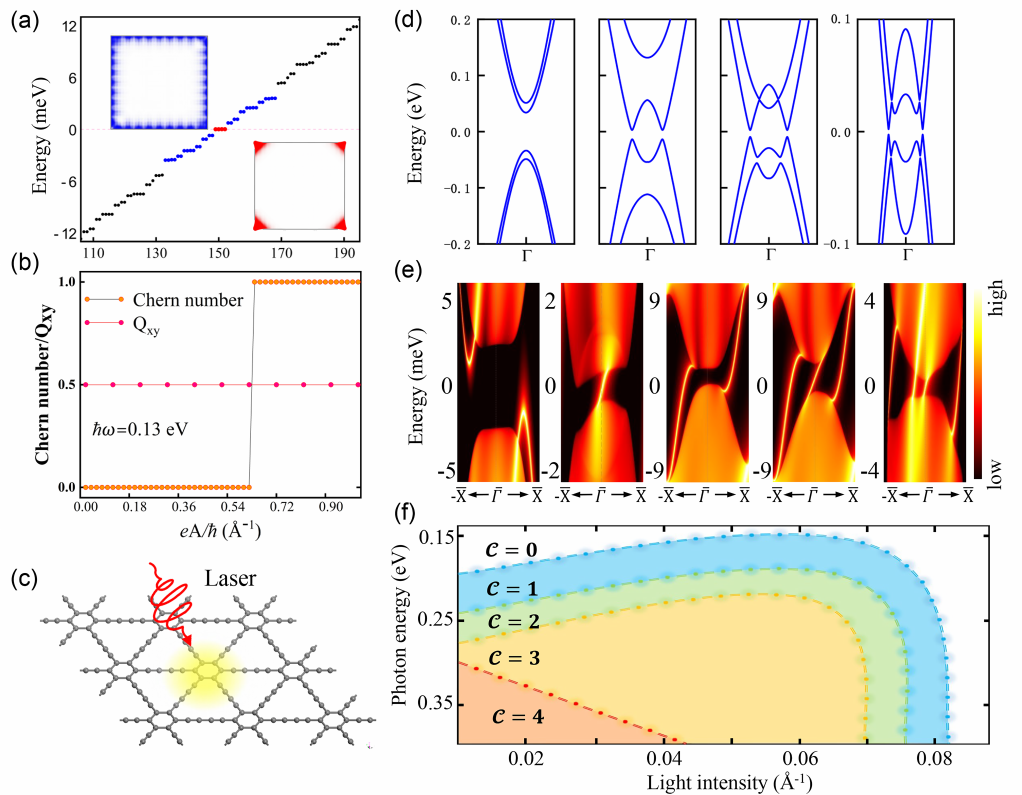}
\caption{Photoinduced high-Chern-number QAH effect from higher-order topological insulators \cite{PhysRevB.109.085148}. (a) The energy spectrum with open boundaries in both the $x$ and $y$ directions, where the edge states and corner states are marked in blue and red, respectively.  The spatial distribution of the corner and edge states are shown as the inserts. (b) The evolution of Chern number $\mathcal{C}$ and quadruple moment $Q_{xy}$ with the increase of light intensity.  (c) Schematic of irradiation of CPL on graphdiyne. (d) The Floquet band structures of graphdiyne around $\Gamma$ under irradiation of CPL with different light intensity $eA/\hbar$ and photon energy $\hbar\omega$. The detailed light intensity and frequency are given in Ref.\cite{PhysRevB.109.085148}. (e) The edge states of graphdiyne along high-symmetry zigzag edge under the irradiation of CPL with different light intensity and photon energy. (f) The Chern number phase diagram of graphdiyne under the irradiation of CPL as a function of light intensity and photon energy.}\label{fig-GDY}
\end{figure*}

Besides, the photoinduced high-Chern-number QAH states can also occur in a higher-order topological insulators. As earlier mentioned, the light-driven BBH model can capture directly the influence of light field on the QTI. Under the irradiation of CPL, the energy spectrum of BBH model shows that both one-dimensional edge states (colored blue) and zero-dimensional corner states (colored red) inside the gap [Fig. \ref{fig-GDY}(a)], indicating the coexistence of QTI and Chern insulator phases. As shown in Fig. \ref{fig-GDY}(b), the quadruple moment $Q_{xy}$ can always be quantized to $1/2$ with the increase of light intensity, indicating that the irradiation of CPL does not destroy the higher-order topology; meanwhile, the Chern number $\mathcal{C}$ accompanies with a transition from 0 to 1. Further, the phase with $\mathcal{C}=2$ or $\mathcal{C}=3$ and  $Q_{xy}=1/2$ are present at higher frequencies~\cite{PhysRevB.109.085148}. 

These photoinduced high-Chern-number QAH states can be realized in experimentally synthesized 2D graphdiyne~\cite{PhysRevLett.123.256402,B922733D}. The crystal structure of graphdiyne constructed from \emph{sp}- and \emph{sp}$^2$-hybridized carbon atoms is shown in Fig.~\ref{fig-GDY}(c). The calculated band structures around $\Gamma$ under various light intensity and photon energy of CPL are shown in Fig. \ref{fig-GDY}(d). 
Similar to the case of BBH model, the multiple band inversion occurs as expected, leading to the Chern number manipulated continuously from trivial state $\mathcal{C}=0$ to nontrivial QAH state with $\mathcal{C}=3$. 
Figure \ref{fig-GDY}(e) illustrates the edge states of graphdiyne under the irradiation of CPL along zigzag direction. One can see that the number of the chiral edge states with different light intensity and  photon energy, further supporting the fact that one can obtain the QAH states by irradiating CPL on a 2D higher-order topological insulators. Moreover, by manipulating the light intensity and  photon energy, one can realize the high-Chern-number QAH states up to $\mathcal{C}=4$ for graphdiyne. The phase diagram can completely summarize the parameter regimes of different phases and gain a deep insight into the topological phase transition. As shown in Fig. \ref{fig-GDY}(f), one can find that there are five distinct phase regimes corresponding to the continuously changed Chern number (ranging from 0 to 4). 

\subsection{Light-manipulated Weyl nodes in topological semimetallic materials}\label{subsec52}

\begin{figure*}
\centering
\includegraphics[width=0.9\textwidth]{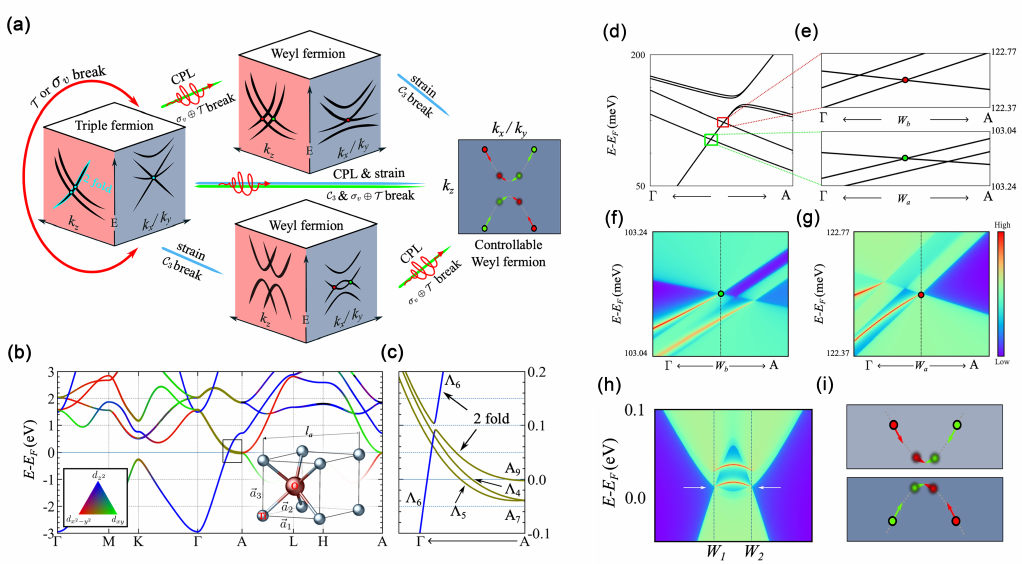}
\caption{Controllable Weyl nodes and fermi arcs from Floquet engineering triple fermions~\cite{huang2024}. (a) The sketched figure illustrated the external field and corresponding symmetry breaking induced topological transition. The dispersion along $k_z$ and $k_x/k_y$ for each phase are illustrated, with the dots in red and green mark different chirality. (b) The electronic band structures of intrinsic WC-type TiO along high-symmetry lines with SOC. The insert depicts the crystal structure of WC-type TiO. (c) The enlarged view of band crossings along $\Gamma$-A. (d) The electronic band structure along $\Gamma$-A under the light intensity of $eA_0/\hbar = 0.100$ \AA $^{-1}$. The enlarged views near Weyl nodes $W_b$ and $W_a$ are shown in (e). (f) and (g) Surface spectrum of (010) surface along Weyl nodes $W_b$ and $W_a$. The projected Weyl nodes with chirality $+1/-1$ are highlighted by red/green circles. (h) The surface spectrum of (010) surfaces along the high-symmetry line connected Weyl nodes with opposite chirality induced by 4\% uniaxial tensile strain. (i) The total 4 pairs of WPs come closer in pairs and vanish under CPL irradiation with increased intensity.}\label{fig-TiO}
\end{figure*}


Distinct from gapped topological phases, such as Chern insulators, topological semimetals possessing the gapless nodal points (or nodal lines) near the Fermi level are particularly relevant to symmetries.
The nontrivial band topology of topological semimetals often leads to attractive phenomena, such as ultrahigh carrier mobility~\cite{shekhar2015extremely,liang2015ultrahigh}, half-integer quantum Hall effects~\cite{novoselov2005two,zhang2005experimental}, large diamagnetism~\cite{ali2014large,chen2016extremely,kumar2017extremely}, and electromagnetic duality~\cite{2017PhysRevB.95.205108,2022PhysRevLett.128.027201}, and thus considered to have a wide range of applications in future devices and technologies.
Benefiting the efforts donated to uncover mappings between symmetries and band topology~\cite{bradlyn2016beyond,YU2022375}, it is now possible to actively manipulate the transition between different topological states and thereby design desired topological semimetals under the irradiation of light fields.
For instance, it has been shown that Floquet WSMs can be generated in materials with QSH state, Dirac fermion, triple fermion, and nodal-line fermion subjected to periodic driving light fields~\cite{PhysRevLett.117.087402,hubener2017,PhysRevB.102.201105}.
Here, we briefly review the material proposal of light-induced WSM from triple fermion in TiO, and nodal-line fermion in carbon allotrope bct-C$_{16}$.

The TiO crystallizing in tungsten carbide type (WC-type) structure with a space group $P\overline{6}m2$ ($D_{3h}$, No.~187) was demonstrated to be an ideal candidate with triple fermions near the Fermi level.
As schematically illustrated in Fig.~\ref{fig-TiO}(a), the triply degenerate nodal points in WC-type TiO without light irradiation are protected by the $C_3$-symmetry, vertical mirror symmetry plus $\mathcal{T}$-symmetry $\mathcal{\sigma}_{v}\oplus\mathcal{T}$. Under time-periodic and space-homogeneous CPL, the light irradiation can break the symmetry of $\mathcal{\sigma}_{v}\oplus\mathcal{T}$ and thereby lead to the triply degenerate nodal points splitting to twofold degenerate Weyl nodes.
The orbital-resolved band structures along high-symmetry paths are shown in Fig. \ref{fig-TiO}(b).
There is a band crossing point along the high-symmetry $\Gamma$-A path, which is mainly contributed by the $e_g$ orbital ($d_{xy}$ and $d_{x^2-y^2}$) and the $d_{z^2}$ orbital.
In fact, the enlarged view of the bands along $\Gamma$-A exhibits three sets of bands with distinct band degeneracy [Fig.~\ref{fig-TiO}(c)], i.e., the non-degenerate $\Lambda _{4}$ and $\Lambda _{5}$ bands, and the doubly degenerate $\Lambda _{6}$ band. The $\Lambda _{4}(\Lambda _{5})$ band crosses with the doubly degenerate $\Lambda _{6}$ band, forming the triply degenerate nodal points that are protected by ${C}_3$ and $\mathcal{\sigma}_{v}\oplus\mathcal{T}$.
The light irradiation can lift this doubly degenerate band, as shown in Figs.~\ref{fig-TiO}(d) and \ref{fig-TiO}(e), forming two non-degenerate bands that are respectively featured by two ${C}_3$ eigenvalues of $e^{i\pi/3}$ and $e^{-i\pi/3}$. Consequently, the triply degenerate nodal points along $\Gamma$-A are absent. These four Weyl points located along the rotation-invariant high-symmetry line $\Gamma$-A are constrained by the rotation symmetry ${C}_3$, which indicates that CPL irradiation breaks $\mathcal{\sigma}_{v}\oplus\mathcal{T}$ but preserve ${C}_3$. The obtained LDOS projected on the semi-infinite (010) surface are plotted in Figs.~\ref{fig-TiO}(f) and ~\ref{fig-TiO}(g), and one can find the characteristic surface states terminated at projections of bulk Weyl nodes.
The application of lattice strain and its coupling of CPL would offer an effective way to manipulate the electronic and topological properties of WC-type TiO. Figure~\ref{fig-TiO}(h) shows the surface states connecting the projections of bulk Weyl nodes are clearly visible under 4\% tensile strain. With various intensities of CPL, the strained WC-type TiO exhibits distinct Weyl semimetallic phases with different numbers of Weyl nodes and Fermi arcs. To be specific, as the light intensity increases, the photon-dressed band structures of strained WC-type TiO show that the two lowest conduction bands and the two highest valence bands simultaneously move away from the Fermi level. Consequently, the Weyl nodes $W_1$ and $W_2$ come together and then annihilate at a specific light intensity as described in Fig.~\ref{fig-TiO}(i).

\begin{figure*}
\centering
\includegraphics[width=0.9\textwidth]{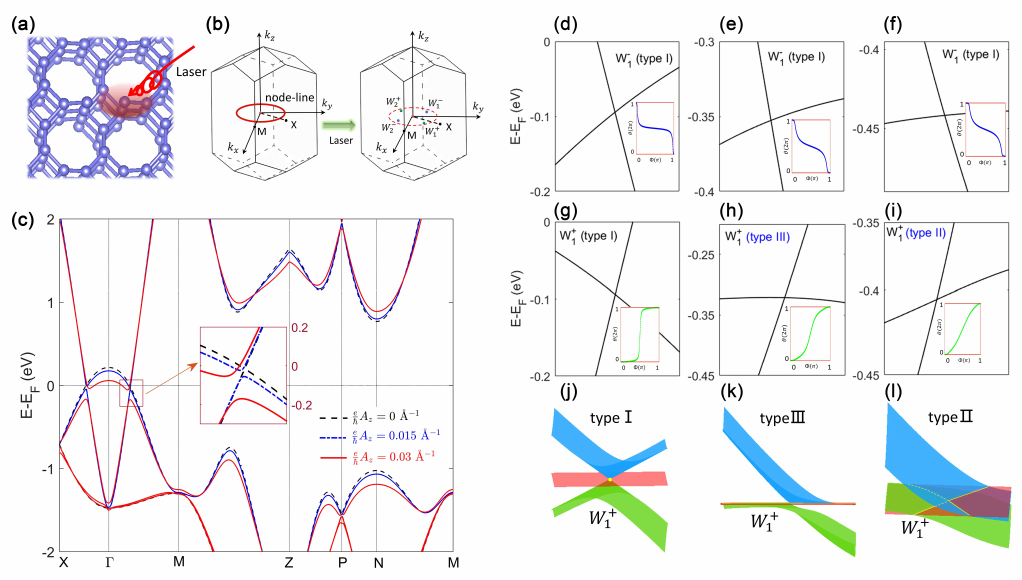}
\caption{Photoinduced Floquet mixed-Weyl semimetallic phase in a carbon allotrope~\cite{PhysRevB.102.201105}. (a) Schematic figure of bct-C$_{16}$ irradiated by incident laser. (b) An ideal Dirac nodal-line located at the $k_x$-$k_y$ plane with $k_z = 0$ is transitioned to two pairs of Weyl nodes once the laser is applied. (c) Floquet band structure evolution of bct-C$_{16}$ under the irradiation of a LPL. Band profiles around one pair of Weyl nodes $W_1^+$ and $W_1^-$ under the irradiation of a LPL with a light intensity (d), (g), and (j) $eA_z /{\hbar}=$0.03 {\AA}$^{-1}$, (e), (h), and (k) $eA_z /{\hbar}=$0.059 {\AA}$^{-1}$, and (f), (i) and (l) $eA_z /{\hbar}=$0.066 {\AA}$^{-1}$. The insets show the evolution of the Wannier charge centers around the Weyl nodes $W_1^-$ and $W_1^+$, respectively.}\label{fig-C16}
\end{figure*}

The carbon allotrope bct-C$_{16}$ crystallizes in a body-centered tetragonal (bct) structure with space group $I4_1/amd$ [Fig. \ref{fig-C16}(a)], which can be obtained from the famous T-carbon through a temperature-driven structural transition \cite{Ding_2020}. Due to the extremely tiny SOC of the carbon element, the interplay between the SOC and light irradiation can be ignored. Therefore, it can be considered as an ideal platform to study the photon-dressed topological states. As shown Fig. \ref{fig-C16}(b), the bct-C$_{16}$ is a NLSMs protected by the $\mathcal{PT}$ symmetry.
The Dirac nodal-line is located at the the mirror reflection invariant $k_x$-$k_y$ plane with $k_z = 0$. Under a periodic field of a linearly polarized laser (LPL), the NLSM phase is transitioned to a mixed-WSM phase with two pairs of tunable Weyl nodes.
As shown in Fig. \ref{fig-C16}(c), one can see that the light irradiation obviously influences the electronic band structures of bct-C$_{16}$. The previous band crossings in the $\Gamma$-$X$ and $\Gamma$-$M$ directions are both gapped, indicating that the nodal-line fermions in bct-C$_{16}$ disappear. With increasing the light intensity, the band gaps are further enlarged [the inset of Fig. \ref{fig-C16}(c)]. The band profiles around one pair of Weyl nodes (i.e., $W_1^+$ and $W_1^-$) evolve with increasing the amplitude of LPL, as shown in Figs. \ref{fig-C16}(d)-(l). The other pair of Weyl nodes shows the same behaviors with respect to the $\mathcal{T}$-symmetry. The band dispersion around the $W_1^-$ with a light intensity $eA_z /{\hbar}$ of 0.03, 0.059, and 0.066 {\AA}$^{-1}$ are illustrated in Figs. \ref{fig-C16}(d), \ref{fig-C16}(e), and \ref{fig-C16}(f), respectively. It is found that the right-handed Weyl node $W_1^-$ keeps to be type-I though it becomes more tilted with increasing the light intensity. On the contrary, light-dependent change around $W_1^+$ is more remarkable. When the light intensity $eA_z /{\hbar}$ increases from 0.03 to 0.066 {\AA}$^{-1}$, the left-handed Weyl node $W_1^+$ undergoes a transition from type-I [Fig. \ref{fig-C16}(g)] to type-II [Fig. \ref{fig-C16}(i)]. In this transition process, the critical type-III Weyl node is present between type-I and type-II states with $eA_z /{\hbar} = 0.059$ {\AA}$^{-1}$ [Fig. \ref{fig-C16}(h)]. The 3D plots of band profiles around $W_1^+$ with $eA_z /{\hbar}$ = 0.03, 0.059, and 0.066 {\AA}$^{-1}$ are respectively shown in Figs. \ref{fig-C16}(j), \ref{fig-C16}(k), and \ref{fig-C16}(l), which are consistent with topologically nontrivial features of type-I, type-II, and type-III Weyl fermions. The results demonstrate that the unconventional Weyl pairs composed of distinct types of Weyl nodes are realized in bct-C$_{16}$.

In addition to the material mentioned above, other realistic material systems with controllable Floquet topological states have also been shown to form the analogous proposals, such as the graphene~\cite{PhysRevB.99.075121}, FeSe~\cite{PhysRevLett.120.156406}, black phosphorus~\cite{PhysRevLett.120.237403}, Na$_3$Bi~\cite{hubener2017}, MnBi$_2$Te$_4$ films~\cite{PhysRevB.107.085151} and so on. These experimentally synthesized materials offer excellent platforms to control the electronic properties for achieving the new and desired topological states.

\section{Summary and perspective}\label{sec6}
Last but not least, we have provided a brief review of recent progress in theoretical investigations of Floquet engineering topological states in condensed-matter systems under light irradiation. The study of various Floquet topological states, such as the light control of type-I, type-II, and type-III Weyl fermions and their topological phase transitions, photoinduced QAH effects with a tunable Chern number, higher-order topological insulators arising from periodic driving and disorder, paves a fascinating path for realizing desired topological states with high tunability. The strong dependence of Floquet topological phases on light-induced symmetry breaking provides a valuable platform for investigating the interplay between symmetry and topology, as well as for generating exotic non-equilibrium topological phases unattainable in equilibrium conditions. Moreover, we show that the combination of first-principles calculations and Floquet theorem offers a reliable avenue to map momentum- and spin- resolved Floquet-Bloch bands in whole Brillouin zone of solids, and most of predicted material candidates are synthesized or well studied in experimental level, facilitating the measurements of Floquet engineering topological states and their device design in experiments. 

Besides, although we point out that the Floquet theorem combining with  effective model as well as first-principles calculations can effectively map  theoretical investigations of Floquet engineering topological states to experiments, but the complex and entangled band manifolds on ultrafast timescales are still not clear. Especially, the thorough understanding of Floquet topological states in periodically driven systems, such as interplay between non-equilibrium topological states and electron–electron corrections, is a long-term challenge.  We believe that further study of Floquet engineering topological states will uncover even more exotic phenomena yet to be imagined, establishing this field as a promising area of study with attractive physical phenomena in the condensed-matter community.

\backmatter

~~\\
\textbf{Acknowledgements}\\
Not applicable.

~\\
\textbf{Funding}\\
This work was supported by the National Natural Science Foundation of China (NSFC, Grants No.~12204074, No.~12222402, No.~92365101, No.~12347101, No.~12304195, No.~12304191, and No.~12074108) and the Natural Science Foundation of Chongqing (Grants No.~2023NSCQ-JQX0024, and No.~CSTB2022NSCQ-MSX0568).

~~\\
\textbf{Availability of data and materials}\\
Not applicable.

~~\\
~~\\
\textbf{\large{Declarations}}

~~\\
\textbf{Ethics approval and consent to participate}\\
Not applicable.

~~\\
\textbf{Consent for publication}\\
Not applicable

~~\\
\textbf{Competing interests}\\
The authors declare no competing interests.

~\\
\textbf{Author contributions}\\
RW and DHX supervised the work. FZ and RC contribute to the work equally. All authors read and edited the full manuscript. Introduction (RW and DHX); basic method of Floquet engineering (RW); light-driven topological semimetallic states from effective models (RC and ZW); Topological states induced by the interplay of periodic driving and disorder (ZN); Floquet engineering of topological states in realistic materials (FZ and DSM); summary (RW). Overview of the review (RW and DHX). All authors have approved the manuscript.


\end{document}